\documentstyle[amssymb,preprint,aps]{revtex}

\begin{document}
\author{S. Bruce\thanks{%
E-mail: sbruce@udec.cl, Fax: 56-41-224520.}}
\title{Comments on the Aharonov-Casher effect}
\address{Departamento de Fisica, Universidad de Concepcion,\\
Casilla 160-C, Concepcion, Chile}
\date{March, 2000}
\maketitle

\begin{abstract}
We study the basic requirements for neutron confinement in the framework of
some 3-D Aharonov-Casher configurations

PACS number(s): 03.65Ge, 03.65.Bz, 12.60.J, 11.30.P.
\end{abstract}

Aharonov and Casher (AC) pointed out the existence of a quantum mechanical
process \cite{AC,K,KA} wherein the behavior of an uncharged dipole is
affected by the presence of an electric field. Let us imagine an
electrically charged object with axial symmetry centered around the $z$
axis. The nearly point dipoles, e.g. neutrons, are completely polarized
along, say, the positive ${\bf y}$ direction. It is straightforward to see
that this system can be recast in a supersymmetric form\cite{RO,BR1,BR2}. To
study supersymmetry breaking, one solves the corresponding eigenvalue
problem for the ground state of the given geometrical configuration.

In this report we shall treat three cases: (1) a finite spherical charge
distribution, (2) an infinite plane with uniform charge density, and (3) the
standard Aharonov-Casher configuration. We conclude that, although in this
circumstance there is apparently no force on the particles, slow neutrons
will tend to move toward regions where the gradient of the field increases.

To start with we assume {\it connectivity} in the configuration space in
order to define a normalizable ground state. The problem turns out to have
exact supersymmetry only under the fulfillment of a condition for the
magnitude of the charge density for the cases of infinite charge
distribution. We also discuss the possibility of {\it supersymmetry} {\it %
breaking} by examining the requirements for the existence of lower energy
bound states.

To be specific, let us consider a spin $1/2$ chargless particle with an
anomalous magnetic moment $\kappa _{n}.$ The Dirac equation can be written 
\cite{H} in a covariant form ($\hbar =c=1$) as 
\begin{equation}
\left( \gamma _{\mu }p^{\mu }-\frac{e\kappa _{n}}{2M_{n}}F^{\mu \nu }\sigma
_{\mu \nu }-M_{n}\right) \Psi (x)=0,  \label{10}
\end{equation}
where $F^{\mu \nu }=\partial ^{\mu }A^{\nu }-\partial ^{\nu }A^{\mu }$ is
the electromagnetic field tensor.

The Aharonov-Casher effective wave equation is obtained by making $A^{0}(%
{\bf x})\neq 0,${\bf \ }${\bf B}({\bf x})={\bf 0,}$ with ${\bf \nabla \cdot
E(x)=}4\pi \rho ({\bf x})$. Equation (\ref{10}) can be recast in the form 
\begin{equation}
\left( {\bf \alpha }\cdot \left( {\bf p}+\frac{ie\kappa _{n}}{M_{n}}\gamma
_{0}{\bf E}({\bf x})\right) +\gamma _{0}M_{n}\right) \Psi ({\bf x,}t)=i\frac{%
\partial }{\partial t}\Psi ({\bf x,}t).  \label{20}
\end{equation}
For stationary states of energy $E$ we write 
\begin{equation}
\Psi _{E}({\bf x,}t)=\Psi _{E}({\bf x})e^{-iEt}=%
{\phi _{E}({\bf x}) \choose \chi _{E}({\bf x})}%
e^{-iEt}.  \label{40}
\end{equation}
Thus from (\ref{20}) and (\ref{40}) we get 
\begin{equation}
\sigma _{k}\otimes \left( p^{k}-\tau _{3}\frac{ie\kappa _{n}}{M_{n}}%
E^{k}\right) \Psi _{E}({\bf x})=\varepsilon \Psi _{E}({\bf x}),  \label{30}
\end{equation}
where $\tau _{3}$ is a $z$-Pauli matrix, ${\bf \sigma =(}\sigma _{1},\sigma
_{2},\sigma _{3}),$ with $\sigma _{i}$ Pauli matrices$,$ $\eta =e\kappa
_{n}/M_{n}$ and $\varepsilon \equiv E^{2}-M_{n}^{2}.$ A $N=1$ supersymmetry
algebra can be constructed in the form 
\begin{equation}
{\cal H}_{SS}{\cal =}\left\{ Q,Q^{\dagger }\right\} ,\qquad \left[ {\cal H}%
_{SS},Q\right] =\left[ {\cal H}_{SS},Q^{\dagger }\right] =0,  \label{60}
\end{equation}
with 
\begin{equation}
{\cal H}_{SS}\Psi _{E}({\bf x})=\frac{\varepsilon }{2M_{n}}\Psi _{E}({\bf x}%
).  \label{70}
\end{equation}
Here 
\begin{equation}
Q\left( {\bf x,p}\right) \equiv \frac{1}{\sqrt{2M_{n}}}\tau ^{-}\otimes {\bf %
\sigma }\cdot \left( {\bf p-}i\eta {\bf E(x)}\right)  \label{80}
\end{equation}
is the {\it supersymmetric charge} and $\tau ^{-}=(1/2)\left( \tau
_{1}-i\tau _{2}\right) ,$ where the $\tau _{1},$ $\tau _{2}$ are Pauli
matrices. Thus ${\cal H}_{SS}$ is invariant under $Q$ and $Q^{\dagger }$.
From (\ref{70}) we find that the equations for $\phi _{E}$ and $\chi _{E}$
are decoupled. In particular, for thermal neutrons we consider the upper
components of $\Psi _{E}$ which satisfy 
\begin{equation}
\left\{ {\bf p}^{2}-\eta \nabla {\bf \cdot E(x)}-\frac{2\eta E{\bf (x)}}{r}%
{\bf \sigma \cdot L}+\eta ^{2}E^{2}({\bf x})\right\} \phi _{E}({\bf x}%
)=\varepsilon \phi _{E}({\bf x}),  \label{90}
\end{equation}
where $r=\left| {\bf x}\right| ,$ with ${\bf L}$ the orbital angular
momentum operator. The supersymmetric generators annihilate the vacuum state
in order to have unbroken symmetry: 
\begin{equation}
Q\phi _{(0)}({\bf x})=0,\qquad Q^{\dagger }\phi _{(0)}({\bf x})=0,
\label{100}
\end{equation}
where $\phi _{(0)}$ is the ground state of the system.

The second equation (\ref{100}) is satisfied identically since in the
nonrelativistic limit the lower components $\Psi _{E=M_{n}}$ vanish. The
first one yields 
\begin{equation}
{\bf \sigma }\cdot \left( {\bf p-}i\eta {\bf E(x)}\right) \phi _{(0)}({\bf x}%
)=0.  \label{130}
\end{equation}
Without lack of generality we can set 
\begin{equation}
\phi _{(0)}({\bf x})=%
{\phi ({\bf x}) \choose 0}%
,\qquad \chi _{(0)}({\bf x})=%
{0 \choose 0}%
.  \label{140}
\end{equation}
Furthermore, in a system with {\it axial} symmetry we have also the
condition $L_{3}\phi _{(0)}({\bf x})=0$, i.e. $\phi _{(0)}({\bf x})=\phi
_{(0)}(r).$ Here then we are concerned with states for which $%
E^{2}=M_{n}^{2} $ ($\varepsilon =0$).

We begin by considering a solid sphere with uniform charge per unit volume $%
\rho _{0}$ centered in the origin of the laboratory frame, so that there
exists an electric field 
\begin{equation}
{\bf E}_{<}{\bf (x)}=4\pi \rho _{0}{\bf x}/3,{\bf \qquad }0{\bf \leq }r\leq
r_{0}\ ;\qquad {\bf E}_{>}{\bf (x)}=4\pi \rho _{0}r_{0}^{3}{\bf x}/3r^{3}%
{\bf ,\qquad }r_{0}\leq r<\infty \ ,  \nonumber
\end{equation}
where $r_{0}$ is the radius of the sphere. In this circumstance there is
apparently no force on the neutrons but there exists a kind of Aharonov-Bohm
effect \cite{AC,K,KA,H}. Nevertheless, if we allow the neutrons to penetrate
the sphere we can consider the problem of the possible bound states of the
neutron in this new AC configuration.

Then from (\ref{130}) we find the first order differential equations 
\begin{equation}
\left( \frac{d}{dr}{\bf -}\beta r\right) \phi _{<}(r)=0,\qquad 0\leq r\leq
r_{0};\qquad \left( \frac{d}{dr}{\bf -}\beta \frac{r_{0}^{3}}{r^{2}}\right)
\phi _{>}(r)=0,\qquad r_{0}\leq r<\infty ,  \label{150}
\end{equation}
where $\beta \equiv 4\pi \rho _{0}\eta /3.$ Thus 
\begin{equation}
\phi _{<}(r)=A_{<}e^{-\beta r^{2}/2},\qquad 0\leq r\leq r_{0};\qquad \phi
_{>}(r)=A_{>}\exp \left( -\frac{\beta r_{0}^{3}}{r}\right) ,\qquad r_{0}\leq
r<\infty ,  \label{160}
\end{equation}
with $A_{\lessgtr }$ complex constants.

Next we demand continuity of the wavefunction and its derivative at $%
r=r_{0}. $ Both conditions yield the same information: 
\begin{equation}
\frac{A_{<}}{A_{>}}=\exp \left( -\frac{1}{2}\beta r_{0}^{2}\right) .
\label{170}
\end{equation}
Moreover, if $\Psi _{E=M_{n}}$ belongs to the Hilbert space, $\phi $ must be
normalizable in ${\Bbb R}^{3}$: 
\begin{equation}
4\pi \lim_{r\rightarrow \infty }\int_{0}^{r}\mid \phi (r^{\prime })\mid
^{2}r^{\prime 2}dr^{\prime }=1.  \label{180}
\end{equation}
However, as $r\rightarrow \infty $ this integral diverges since $\exp \left(
-\beta r_{0}^{3}/r\right) \rightarrow 1.$ Therefore supersymmetry is broken
in the case.

Next we solve the general problem (\ref{90}) by separation of variables: $%
\phi ({\bf x})=\phi (r){\cal Y}_{l,j,m}\left( \theta ,\varphi \right) $,
where in terms of the spherical harmonics $Y_{lm_{l}},$ 
\begin{equation}
{\cal Y}_{l,l\pm 1/2,m_{j}}\left( \theta ,\varphi \right) =\frac{1}{\sqrt{%
2l+1}}\left\{ \pm \sqrt{l\pm m_{j}+\frac{1}{2}}Y_{lm_{j}-1/2}\left( \theta
,\varphi \right) \left( 
\begin{array}{c}
1 \\ 
0
\end{array}
\right) \right.  \label{210}
\end{equation}
\[
+\left. \sqrt{l\mp m_{j}+\frac{1}{2}}Y_{lm_{j}+1/2}\left( \theta ,\varphi
\right) \left( 
\begin{array}{c}
0 \\ 
1
\end{array}
\right) \right\} . 
\]
Thus from (\ref{90}) and (\ref{210}) we get 
\begin{eqnarray}
\left( \frac{d^{2}}{dr^{2}}-\frac{l\left( l+1\right) }{r^{2}}+\varepsilon
-2\beta \left( {\bf \sigma \cdot L}-\frac{3}{2}\right) -\beta
^{2}r^{2}\right) \psi _{<}(r) &=&0,\qquad r\leq r_{0},  \label{220} \\
\left( \frac{d^{2}}{dr^{2}}-\frac{l\left( l+1\right) }{r^{2}}+\varepsilon 
{\bf -}2\beta \left( \frac{r_{0}^{3}}{r^{3}}\right) {\bf \sigma \cdot L}%
-\beta ^{2}\left( \frac{r_{0}^{3}}{r^{2}}\right) ^{2}\right) \psi _{>}(r)
&=&0,\qquad r>r_{0},  \label{230}
\end{eqnarray}
where $\psi _{\lessgtr }(r)\equiv r\phi _{\lessgtr }(r).$ The radial
solutions must be normalizable in the range $0\leq r<\infty $, and we also
demand continuity at $r_{0}$ on the corresponding solutions$.$ For $\psi
_{<}(r)$ we find 
\begin{equation}
\left( \frac{d^{2}}{dr^{2}}-\frac{l\left( l+1\right) }{r^{2}}+\epsilon _{\pm
j}-\beta ^{2}r^{2}\right) \psi _{<}(r)=0,  \label{240}
\end{equation}
with $\epsilon _{\pm j}\equiv \varepsilon +\beta \left( 3\mp 2\left(
j-1/2\right) \right) $. Thus 
\begin{equation}
\psi _{<}(r)=C\,_{1}F_{1}\left( \frac{l+3/2-\epsilon _{\pm j}}{2}%
;l+3/2+1;\beta r^{2}\right) r^{l+1}e^{-\beta r^{2}/2},  \label{250}
\end{equation}
where $C$ is a complex constant and $_{1}F_{1}$ is the {\it confluent
hypergeometric function}.

For $r\geq r_{0}$ we obtain 
\begin{equation}
\left( \frac{d^{2}}{dr^{2}}-\frac{l\left( l+1\right) }{r^{2}}+\varepsilon 
{\bf \mp }2\beta r_{0}^{3}\frac{j-1/2}{r^{3}}-\beta ^{2}\left( \frac{%
r_{0}^{3}}{r^{2}}\right) ^{2}\right) \psi _{>}(r)=0.  \label{260}
\end{equation}
Eq.(\ref{260}) has only one kind of solution: non-normalizable
scattering-like states for $\varepsilon >0$ ($E^{2}>M_{n}^{2}$ ) since the
potential decreases as $V\sim 1/r^{2}$ ($1/r^{3}\sim dV/dr$), and also
because there is a term proportional to $1/r^{4}$ induced by the electric
moment of the particle\cite{AC}.

The second case considers an infinite plane of thickness $L$ with uniform
charge density $\rho $, situated symmetrically on the $xy$ plane. This
configuration resembles a potential well in one-dimensional quantum
mechanics. The electric field is given by 
\begin{equation}
{\bf E}_{<}{\bf (}z{\bf )}=4\pi \rho _{0}z\widehat{k},{\bf \qquad }\left|
z\right| \leq \frac{L\ }{2};\qquad {\bf E}_{>}{\bf (}z{\bf )}=4\pi \rho _{0}L%
\frac{z}{\left| z\right| }\widehat{k}{\bf ,\qquad }\left| z\right| >\frac{L\ 
}{2}\ .  \nonumber
\end{equation}
We assume that the neutrons are completely polarized along the positive $%
{\bf z}$ direction. Thus 
\begin{equation}
\phi ({\bf x})=R\left( r\right) \phi (z)\exp \left( \pm i\nu \varphi \right)
.  \label{270}
\end{equation}
In order that the wavefunction be single valued when the full azimuth is
allowed, $\nu $ must be an integer. Therefore from (\ref{90}) and (\ref{270}%
) we obtain the differential equations 
\begin{eqnarray}
\frac{1}{R\left( r\right) }\frac{1}{r}\frac{d}{dr}\left( r\frac{dR\left(
r\right) }{dr}\right) -\frac{\nu ^{2}}{r^{2}}+\frac{1}{\phi (z)}\frac{%
d^{2}\phi (z)}{dz^{2}}{\bf +}4\pi \eta \rho _{0}-16\pi ^{2}\rho _{0}^{2}\eta
^{2}z^{2}+\varepsilon  &=&0,{\bf \qquad }\left| z\right| \leq \frac{L\ }{2},
\\
\frac{1}{R\left( r\right) }\frac{1}{r}\frac{d}{dr}\left( r\frac{dR\left(
r\right) }{dr}\right) -\frac{\nu ^{2}}{r^{2}}+\frac{1}{\phi (z)}\frac{%
d^{2}\phi (z)}{dz^{2}}+\varepsilon  &=&0,{\bf \qquad }\left| z\right| >\frac{%
L\ }{2}.
\end{eqnarray}
Thus 
\begin{equation}
\left( \frac{1}{r}\frac{d}{dr}\left( r\frac{d}{dr}\right) +\left( k^{\prime
2}-\frac{\nu ^{2}}{r^{2}}\right) \right) R_{\nu k^{\prime }}\left( r\right)
=0
\end{equation}
is the radial equation for a Bessel function, with $R_{\nu k^{\prime
}}\left( r\right) =CJ_{\nu }\left( k^{\prime }r\right) $, where $k^{\prime
2}=k^{2}-\varepsilon $, with $k$ a real positive parameter. Hence for the
ground state ($\varepsilon =0$) we have 
\begin{eqnarray}
\left( \frac{d^{2}}{dz^{2}}+4\pi \eta \rho _{0}-k^{2}-16\pi ^{2}\rho
_{0}^{2}\eta ^{2}z^{2}\right) \phi (z) &=&0,\qquad L/2\geq \left| z\right| ,
\\
\left( \frac{d^{2}}{dz^{2}}-k^{2}\right) \phi (z) &=&0,\qquad L/2<\left|
z\right| .
\end{eqnarray}
The (normalizable) ground state is then 
\begin{equation}
\phi _{k}(z)=\left\{ 
\begin{array}{c}
A\exp \left( -\frac{1}{2}k^{2}z^{2}\right) ,\qquad \qquad \ \qquad \quad
L/2\geq \left| z\right| , \\ 
A\exp \left( -\frac{1}{2}k^{2}L^{2}+k\left( L-z\right) \right) ,\qquad
L/2<\left| z\right| ,
\end{array}
\right. 
\end{equation}
where the values of $k$ are restricted by the condition $4\pi \eta \rho
_{0}>k^{2}.$ Note that $k$ denotes an infinite degeneracy in the ground
state which proceeds from the unbound motion of the particle on the $xy$
plane. As an example we insert $c$ into the previous inequality and get $%
4\pi e\kappa _{n}\rho _{0}/\left( M_{n}c^{2}\right) >k^{2}.$ Choosing $\rho
_{0}=$ $2.0\times 10^{6}$ [esu/cm$^{3}$] we find the bound $15.\,28$ $[$cm$%
^{-1}]>k.$ As one would expect, this result does not depend on the plane
thickness $L.$

Finally let us examine the standard 1+2 AC configuration\cite{BR1,BR2}. The
problem turns out to have exact supersymmetry only under the fulfillment of
a condition for the magnitude of the charge distribution which generates the
electric field.

An infinite cylinder with uniform charge per unit volume $\rho $ centered
along the $z$ axis, generates the electric field 
\begin{equation}
{\bf E}_{<}{\bf (x)}=\rho {\bf x}/2,{\bf \qquad }0{\bf \leq }r\leq
r_{0};\qquad {\bf E}_{>}{\bf (x)}=\rho r_{0}^{2}{\bf x}/2r^{2}{\bf ,\qquad }%
r_{0}\leq r<\infty ,  \label{300}
\end{equation}
where $r_{0}$ is the radius of the cylinder and for simplicity we have
chosen $\widehat{{\bf x}}{\bf \cdot \widehat{z}=}0.$ Here $\widehat{{\bf x}}$
and $\widehat{{\bf z}}$ are unit vectors in the ${\bf x}$ and ${\bf z}$
directions respectively. The neutrons are completely polarized along the
positive ${\bf z}$ direction. They move on a plane in the presence of ${\bf %
E.}$

Then again from (\ref{130}) we find the differential equations 
\begin{equation}
\left( \frac{d}{dr}{\bf -}\beta r\right) \phi _{<}(r)=0,\qquad 0\leq r\leq
r_{0};\qquad \left( \frac{d}{dr}{\bf -}\frac{\beta r_{0}^{2}}{r}\right) \phi
_{>}(r)=0,\qquad r_{0}\leq r<\infty ,  \label{420}
\end{equation}
where $\beta \equiv -e\rho \kappa _{n}/4M_{n}.$ Thus 
\begin{equation}
\phi _{<}(r)=Ae^{\frac{1}{2}\beta r^{2}},\qquad 0\leq r\leq r_{0};\qquad
\phi _{>}(r)=Br^{\beta r_{0}^{2}},\qquad r_{0}\leq r<\infty ,  \label{430}
\end{equation}
with $A,B$ complex constants.

Next we demand continuity of the wavefunction and its derivative at $r=r_{0}$
yielding the boundary condition $A\exp \left[ \left( 1/2\right) \beta
r_{0}^{2}\right] =Br_{0}^{\beta r_{0}^{2}}$. Furthermore, if $\Psi _{E=M_{n}}
$ belongs to the Hilbert space, $\phi $ must be normalizable on the plane $%
[0,2\pi ]\times \lbrack 0,\infty \lbrack $ and thus we must require that $%
\beta r_{0}^{2}<-1.$ This inequality constitutes a necessary requirement on
the possible values of $\lambda \equiv \rho \pi r_{0}^{2}$ if we want to
preserve unbroken supersymmetry. As $\lambda $ depends linearly on $r_{0}^{2}
$, one can in principle set up a configuration with the required $\lambda $ 
\cite{AC,BR1}$.$ For instance, cnserting $c$ into the expression for $%
\lambda ,$ we get $\left| \lambda \right| _{\min }\backsimeq 4\pi
M_{n}c^{2}/\left| e\kappa _{n}\right| \backsimeq 60.\,62$ $\times 10^{6}$
[esu/cm]. Naturally, this result is independent of the charged line diameter 
$2r_{0}$.

To treat the general eigenvalue problem, we observe that the eigenvalue
problem stated by (\ref{90}) has two kinds of solutions\cite{BR1}: a)
non-normalizable scattering-like states for $\varepsilon >0$ $\left(
E^{2}>M_{n}^{2}\right) $; b) normalizable bound states for $\varepsilon <0$ $%
\left( E^{2}<M_{n}^{2}\right) $ . The energy levels are obtained by
requiring that the radial solutions and their derivatives be continuous at $%
r=r_{0},$ i.e., this is the quantization condition for the remaining energy
levels.

From the above we can draw the conclusion that electric charge distribution
has to be sufficiently spread out in space in order to preserve unbroken
supersymmetry. If this be the case, $\phi $ is normalizable and thus $\Psi
_{E=M_{n}}$ constitutes a bound state of the system. Furthermore the
magnitude of the electric charge distribution has to be sufficiently large
in order to generate a bound states. Confinement is usually achieved by
means of diverse magnetic traps. Cold neutrons are extensively used to test
quantum theory, and in applied physics \cite{TR,AG,EX}.

{\bf Acknowledgments}

This work was supported by Direcci\'{o}n de Investigaci\'{o}n, Universidad
de Concepci\'{o}n, through grants P.I. 99.11.27-1.0.

I am grateful to the School of Physics, University of Melbourne, Australia.
I am very thankful to Professors A. G. Klein and G. I. Opat for their
valuable criticisms and helpful suggestions on the various aspects of this
paper.

\end{document}